\documentclass[showpacs, preprint, floatfix]{revtex4-1}

\usepackage{amsmath}
\usepackage{graphicx}
\usepackage{textcomp}
\usepackage{url}
\begin{document}

\title{ Current density operator in systems with non-parabolic, position-dependent 
energy bands }

\author{William R. Frensley}
\affiliation{Electrical Engineering, University of Texas at Dallas, Richardson, TX 75080}
\email{frensley@utdallas.edu}

\begin{abstract}
The present manuscript was written in 1994 and was not published.  It addresses the form that the quantum-mechanical
current density must take in mesoscopic treatments of semiconductor heterostructures, in which the electron dispersion
relations are non-parabolic and position dependent, rendering the textbook expressions inapplicable.  The approach is
to derive the continuity equation for the specific model under consideration, using generalizations of Green's 
identities to higher-order derivatives and to discrete models of different topological structure.  A new addendum addresses
two issues of more current interest: the use of irregular meshes in discrete formulations, and the identification of the 
Heisenberg velocity operator to evaluate current density.  It is demonstrated that on discrete domains the velocity operator
fails to satisfy a sensible continuity equation, and therefore cannot be identified with the current density.
\end{abstract}


\maketitle

\section*{Original 1994 Abstract}

 In semiconductor heterostructures the electron dispersion relations are non-parabolic
and vary with position.  The non-parabolicity is described by effective Hamiltonians 
which either include higher-order 
derivatives or couple several basis states.  
As a result, the current density operator is not simply 
related to the gradient.  By generalizing Green's identity to higher-order derivatives 
and to difference relations, the appropriate form of this operator is derived for all of 
the commonly-used band structure representations.

\section{Introduction}

  The wave mechanics of semiconductor heterostructures is complicated by the fact that
the electron dispersion relation (or energy band structure) is generally non-parabolic
at modest energies and necessarily varies with position.  Under such circumstances, 
the form of the current density operator is no longer simply a symmetrized gradient.

  The form of the particle current density operator $J$ is clearly constrained by the 
group-velocity theorem \cite{kroemer75}, 
so that the expectation value of $J$ on a state of definite wavevector $k$ is
\begin{equation}  \label{eqn:group_vel}
 \langle k | J | k \rangle = v_g \langle k | k \rangle 
 = \partial E / \partial k \langle k | k \rangle / \hbar.
\end{equation}
This equation, together with the band energies and eigenstates, in principle determines
the form of $J$.  It is, however, much more convenient to directly derive $J$ from 
the Hamiltonian for a given problem.  One does so by evaluating the time derivative 
of the probability density:
\begin{equation}  \label{eqn:drhodt}
 \partial \psi^* \psi / \partial t = (1/i\hbar) [ \psi^* (H \psi) - (H \psi^*) \psi ].
\end{equation}
Green's identity 
[$f \nabla^2 g - g \nabla^2 f = \nabla \cdot (f \nabla g - g \nabla f)$],
or a generalization thereof, 
is then invoked to write the right-hand side of (\ref{eqn:drhodt})
as the divergence of the current density.

   Heterostructures are most often described at a ``mesoscopic'' level where the
microscopic (smaller than the atomic diameter) behavior of the wavefunction can be 
factored out.  In order to realistically describe the  non-parabolic dispersion 
relation, the resulting effective Hamiltonian must be more elaborate than a 
simple Laplace operator, and
Green's identity must be correspondingly be generalized to derive the form of $J$.  
The commonly-used mesoscopic models can be classified as effective-mass or tight-binding
approaches.

\section{ The Current Density in Specific Representations }

   For the purposes of the present discussion, we will assume that we possess a 
hermitian effective Hamiltonian which is valid throughout the structure, including 
any interval containing an abrupt heterointerface.  The issues thus presumed to have
been resolved have traditionally been posed as the definition of matching conditions
for the mesoscopic wavefunction.  These matching conditions are frequently 
``derived'' from the continuity of $J$
\cite{harrison61,bendaniel66,ando82}, 
but this condition is not sufficient to 
uniquely determine them \cite{zhu_kroemer83}.  
The Hamiltonian and the matching conditions are equivalent
pieces of information, in the sense that either one may be derived from the other.
In contrast, the continuity of $J$ follows solely from the hermiticity of $H$ 
\cite{hj_note}. 

   For the sake of simplicity, only the current density in one dimension will be 
considered.  Extension to three-dimensional structures is in all cases straightforward,
if somewhat tedious.

\subsection{ Wannier-Slater Effective Mass Theory }

  In the approach to effective-mass theory proposed by Wannier \cite{wannier37}
and expounded by Slater \cite{slater49},
the microscopic wavefunction $\psi$ is expanded as a linear combination of 
localized Wannier functions, each of which is centered within a different unit cell.  
The expansion coefficients $\Psi_i$ can be regarded as values of a discrete lattice 
function which is interpolated between lattice points by a continuous function 
$\Psi (z)$, for which an effective-mass Schroedinger equation is derived.  If the
electron dispersion relation can be expanded as
\begin{equation}
 E(k) = \sum_{n=0}^{N} A_n(z) k^{2n},
\end{equation}
then the effective Hamiltonian is 
\begin{equation}  \label{eqn:wannierH}
 H^{\rm (WS)} \Psi =  \sum_{n=0}^{N} (-1)^n \frac{\partial^n}{\partial z^n} A_n (z)
 \frac{\partial^n \Psi}{\partial z^n} + V (z) \Psi.
\end{equation}
To derive the appropriate $J$ for this Hamiltonian, we require a generalized Green's
identity:  \widetext
\begin{equation}  \label{eqn:genGreen}
  f \frac{\partial^n}{\partial z^n} A \frac{\partial^n g}{\partial z^n} 
   - g \frac{\partial^n}{\partial z^n} A \frac{\partial^n f}{\partial z^n} 
= \frac{\partial}{\partial z} \sum_{j=0}^{n-1} (-1)^j \left[
  \frac{\partial^j f}{\partial z^j} \frac{\partial^{n-j-1}}{\partial z^{n-j-1}} A
  \frac{\partial^n g}{\partial z^n} - 
  \frac{\partial^j g}{\partial z^j} \frac{\partial^{n-j-1}}{\partial z^{n-j-1}} A
  \frac{\partial^n f}{\partial z^n} \right] .
\end{equation}
(This identity is readily proven by expanding the derivative on the right-hand side;
the summation then becomes a telescoping series.)  The value of current density is
thus 
\begin{equation}  \label{eqn:wannierJ}
  \langle J^{\rm (WS)}_z \rangle (z) = \frac{i}{\hbar} \sum_{n=1}^{N} \sum_{j=0}^{n-1} 
    (-1)^{n-j} 
  \left[ \frac{\partial^j \Psi^*}{\partial z^j} 
    \frac{\partial^{n-j-1}}{\partial z^{n-j-1}} A_n \frac{\partial^n \Psi}{\partial z^n}
- \frac{\partial^j \Psi}{\partial z^j} 
    \frac{\partial^{n-j-1}}{\partial z^{n-j-1}} A_n \frac{\partial^n \Psi^*}{\partial z^n}
\right].
\end{equation}  
Some formulations of the matching conditions produce terms in the Hamiltonian 
of the form 
\begin{equation}  \label{eqn:odd_order}
H^{(2n+1)} = \frac{i}{2} (B_{2n+1} \partial^{2n+1} / \partial z^{2n+1} 
+ \partial^{2n+1} / \partial z^{2n+1} B_{2n+1}).  
\end{equation}
Their contribution to the current density
may be readily derived from another identity:
\begin{equation}
    f \frac{\partial^{2n+1} g}{\partial z^{2n+1}} 
  + g \frac{\partial^{2n+1} f}{\partial z^{2n+1}}
 = \frac{\partial}{\partial z} \sum_{j = 0}^{2n} (-1)^j 
   \frac{\partial^j f}{\partial z^j} 
  \frac{\partial^{2n-j} g}{\partial z^{2n-j}},
\end{equation}
leading to contributions to the current density of
\begin{equation}    \label{eqn:Jodd_order}
 \langle J^{(2n+1)} \rangle (z) = -\frac{1}{2\hbar} \sum_{j=0}^{2n} (-1)^j
 \left[ \frac{\partial^j B_{2n+1} \psi^*}{\partial z^j} 
    \frac{\partial^{2n-j} \psi}{\partial z^{2n-j}}
  + \frac{ \partial^j \psi^*}{\partial z^j} 
     \frac{\partial^{2n-j} B_{2n+1} \psi}{\partial z^{2n-j}} \right].
\end{equation}
It appears (based upon an examination of some low-order cases) that any apparently
hermitian differential operator \cite{hermiticity_note} can be manipulated into an
expression containing only terms of the form (\ref{eqn:wannierH}) and 
(\ref{eqn:odd_order}).  

\subsection{ Luttinger-Kohn Effective Mass Theory }

   The Luttinger-Kohn approach to effective-mass theory \cite{kohnlutt54,bastard88,burt87a} 
more conveniently includes
the effects of several bands.  In this scheme the microscopic 
wavefunction is decomposed into the ${\bf k} = 0$ Bloch functions and a set of 
slowly-varying envelope functions $\chi_m (z)$, $m$ being a band index.  Differing 
numbers of bands may be included, with perhaps the most general scheme being that 
 derived by Bastard \cite{bastard88}. 
By regrouping the various terms of Bastard's Hamiltonian with respect to the 
derivatives,
 it can be written in the form \cite{bastard_note} 
(summations implied over repeated indices)
\begin{equation}  \label{eqn:kohnluttH}
  (H^{\rm (LK)} \chi)_l 
  = \left[ -\frac{\partial}{\partial z} A_{lm} \frac{\partial}{\partial z}
  - \frac{i}{2} (B_{lm} \frac{\partial}{\partial z} + \frac{\partial}{\partial z} 
   {B^\dagger}_{lm}) 
 + C_{lm} \right] \chi_m,
\end{equation}
where $A$ and $C$ are hermitian matrices and $B$ a general matrix.  $A$, $B$, and $C$ are
 indexed by the band label $m$, and are 
$z$-dependent in a heterostructure.  
Using the above identities, the current density is readily shown to be
\begin{equation}  \label{eqn:kohnluttJ}
  \langle J^{\rm (LK)}_z \rangle (z) = \frac{1}{i\hbar} \left[
   \chi_l^* A_{lm} \frac{\partial \chi_m}{\partial z} 
   - \frac{\partial \chi^*_l}{\partial z} A_{lm} \chi_m 
   + \frac{i}{2} \chi^*_l (B + B^\dagger)_{lm} \chi_m \right].
\end{equation}  
Similar expressions have been obtained by Altarelli \cite{altarelli83a} and by Burt
\cite{burt87a}.

\subsection{ Tight-Binding Theories }

  In the tight-binding approach \cite{slater_koster54}, 
the wavefunction is expanded in terms of a set of 
localized states $m = 1, \ldots, M$ in each atomic layer $j$
\begin{equation}
  | \psi \rangle = \sum_{j,m} c_{jm} | j m \rangle.
\end{equation} 
The coefficients $c_{jm}$ can be thought of as forming a block-structured vector ${\bf c}$
with vector elements ${\bf c}_j = [c_{j1}, \ldots, c_{jM}]^T$.  The Hamiltonian then
becomes a block-structured matrix ${\bf H}^{\rm (TB)}$, of which the diagonal blocks 
${\bf H}^{\rm (TB)}_{ii}$ are hermitian and
describe interactions within a plane and the off-diagonal blocks ${\bf H}^{\rm (TB)}_{ij}$
are not necessarily hermitian 
(${\bf H}^{\rm (TB)}_{ij} = {\bf H}^{{\rm (TB)} \dagger}_{ji}$)
and describe the coupling between planes.  If only nearest-neighbor interactions are
included, ${\bf H}^{\rm (TB)}$ is block-tridiagonal \cite{schulman79,ting92a}
(${\bf H}^{\rm (TB)}_{ij} \ne 0$ only for $j = i-1, i, i+1$).

  Because the tight-binding representation is intrinsically discrete, we need to
modify somewhat our concepts of probability and current density.  A total probability 
density $\rho_i$ is associated with each atomic plane $i$, and is equal to 
${\bf c}_i^\dagger {\bf c}_i$.  The current density represents the flux between 
adjacent planes; we will write the flux between planes $i$ and $i+1$ as 
$J_{i+1/2}$ \cite{discr_note}.
Applying (\ref{eqn:drhodt}) to the tight-binding Hamiltonian ${\bf H}^{\rm (TB)}$ and assuming only nearest-neighbor interactions, we get  \widetext
\begin{equation}
  \frac{\partial}{\partial t} {\bf c}_i^\dagger {\bf c}_i = \frac{1}{i\hbar} \left[
  {\bf c}_{i}^\dagger {\bf H}^{\rm (TB)}_{i,i-1} {\bf c}_{i-1}
 + {\bf c}_{i}^\dagger {\bf H}^{\rm (TB)}_{i,i+1} {\bf c}_{i+1}
 - {\bf c}_{i-1}^\dagger {\bf H}^{\rm (TB)}_{i-1,i} {\bf c}_{i}
 - {\bf c}_{i+1}^\dagger {\bf H}^{\rm (TB)}_{i+1,i} {\bf c}_{i}  \right].
\end{equation}
This can be written as a discrete continuity equation,
\begin{equation}  \label{eqn:discrContinuity1}
  \partial \rho^{\rm (TB)}_i / \partial t = \langle J^{\rm (TB)} \rangle_{i-1/2} 
   - \langle J^{\rm (TB)} \rangle_{i+1/2},
\end{equation}
if $\langle J^{\rm (TB)} \rangle$ is identified as
\begin{equation}
  \langle J^{\rm (TB)} \rangle_{i+1/2} = \frac{1}{i \hbar} \left( 
     {\bf c}_{i+1}^\dagger {\bf H}^{\rm (TB)}_{i+1,i} {\bf c}_i 
     - {\bf c}_{i}^\dagger {\bf H}^{\rm (TB)}_{i,i+1} {\bf c}_{i+1}
    \right).
\end{equation} 
Here we see the machinery of Green's identity operating in a discrete space.

   A variation of the tight-binding scheme is the ``Wannier Orbital Model'' 
\cite{ting87a}, which draws upon the discrete form of the Wannier-Slater theory.  
Interactions with remote neighbors are included to fit the $E(k)$ dispersion, and
typically $M=1$, so that only one band is modeled.  Inserting such a Hamiltonian
into (\ref{eqn:drhodt}) leads to many terms, which cannot be associated with 
a particular position.  Thus, the notion
of a local current density disappears, due to the direct interactions between remote
sites.  Instead, we may define an antisymmetric current matrix with elements 
\begin{equation}
  \langle J^{\rm (WO)} \rangle_{ij} = (i/\hbar) 
   \left( c_i^* H^{\rm (WO)}_{ij} c_j - c_j^* H^{\rm (WO)}_{ji} c_i  \right),
\end{equation}
where $\langle J \rangle_{ij}$ is the current flowing out of site $i$ into site $j$.
The nonlocal continuity equation is then
\begin{equation}
  \partial \rho^{\rm (WO)}_i / \partial t = -\sum_j \langle J^{\rm (WO)} \rangle_{ij}.
\end{equation}

\section{ Summary}

   The form of the current density has been derived for all of the usual models of 
heterostructure electronic states.  While one can usually find a way to obtain 
correct answers in a manual calculation without knowledge of the general expressions
for $J$ presented here, they provide a useful check on the results.  
The use of these expressions becomes more necessary if one 
seeks to develop the machinery for automatic computations (numerical or symbolic)
applicable to a wide variety of heterostructures.  

  The reader will have noticed that, in conformance to the practice in quantum-mechanics
texts, expressions for the expectation values of $J$ have been presented, not 
expressions for the operator itself.  It is very difficult to represent $J$ as an
ordinary quantum-mechanical operator.  
Actually, $J$ is much more naturally expressed
as a superoperator which acts upon a density operator 
\cite{frensley90a}.  All of the results presented here may be derived much more
elegantly in a superoperator formalism, at the cost of an unfamiliar notation.

\section*{References}

\bibitem{kroemer75} H.~Kroemer, Proc.\ IEEE {\bf 63}, 988 (1975).

\bibitem{harrison61} W.~A.~Harrison, Phys.\ Rev.\ {\bf 123}, 85 (1961).

\bibitem{bendaniel66} D.~J.~BenDaniel and C.~B.~Duke, Phys.\ Rev.\ {\bf 152}, 683 (1966).

\bibitem{ando82} T.~Ando and S.~Mori, Surf.\ Sci.\ {\bf 113}, 124 (1982).

\bibitem{zhu_kroemer83} Q.-G.~Zhu and H.~Kroemer, Phys.\ Rev.\ B {\bf 27}, 3519 (1983).

\bibitem{hj_note} The form of the effective Hamiltonian cannot be completely 
determined at the mesoscopic level.  One requires a realistic microscopic model
from which the mesoscopic approximation may be derived.  Nevertheless, the 
commonly used forms, obtained by placing material-dependent quantities in the 
center of the derivatives or taking average values for inter-atomic-plane matrix
elements, appear to be adequate for many systems. 

\bibitem{wannier37} G.~H.~Wannier,  Phys.\ Rev.\ {\bf 52}, 191 (1937).

\bibitem{slater49} J.~C.~Slater, Phys.\ Rev.\ {\bf 76}, 1592 (1949).

\bibitem{hermiticity_note} By ``apparently hermitian,'' I mean an operator in which the
$z$-dependent parameters appear symmetrically with respect to the derivatives.  The 
rigorous demonstration of hermiticity requires a specification of the boundary conditions
applied to the operator.  For an extensive discussion of the physical consequences of 
boundary conditions, see \cite{frensley90a}.

\bibitem{kohnlutt54} J.~M.~Luttinger and W.~Kohn, Phys.\ Rev.\ {\bf 97}, 
869 (1955).

\bibitem{bastard88} G.~Bastard, {\it Wave Mechanics Applied to Semiconductor 
Heterostructures} (Halsted Press, New York, 1989), ch.~3.

\bibitem{burt87a} M.~G.~Burt, Semicond.\ Sci.\ Technol.\ {\bf 2}, 460 (1987).

\bibitem{bastard_note} Bastard's Hamiltonian [equation (20) of chapter 3] contains
a term of the form $i \langle l | p_z | m \rangle \partial / \partial z$, which is
hermitian only if this matrix element is independent of $z$. Burt \cite{burt87a}  
has pointed out that a rigorous derivation of the envelope-function approach requires
that the same Bloch functions be used throughout the heterostructure, which would assure
the hermiticity.  In practice this condition is often violated as the Bloch functions
are presumed to vary with the local material composition.

\bibitem{altarelli83a} M.~Altarelli, Phys.\ Rev.\ B {\bf 28}, 842 (1983).

\bibitem{slater_koster54} J.~C.~Slater and G.~F.~Koster, Phys.\ Rev.\ {\bf 94}, 
  1498 (1954).

\bibitem{schulman79} J.~N.~Schulman and T.~C.~McGill, Phys.\ Rev.\ B {\bf 19},
6341 (1979).

\bibitem{ting92a} D.~Z.-Y.~Ting, E.~T.~Yu, and T.~C.~McGill, Phys.\ Rev.\ B {\bf 45},
3583 (1992).

\bibitem{discr_note} A similar concept and notation for the current density appears
in finite-difference numerical computations.  See, for example, S.~Selberherr,
{\it Analysis and Simulation of Semiconductor Devices} (Springer, Vienna, 1984),
ch.\ 6.

\bibitem{ting87a} D.~Z.-Y.~Ting and Y.-C.~Chang, Phys.\ Rev.\ B {\bf 36}, 4359 (1987).  

\bibitem{frensley90a} W.~R.~Frensley, Rev.\ Mod.\ Phys.\ {\bf 62}, 745 (1990).

\section{2015 Addendum}

\subsection{Irregular Triangular Meshes}

The finite-element technique has become a popular way of solving problems that are described by partial differential
equations.  One of the chief characteristics of this technique is the use of geometrically and topologically irregular
meshes.  Discussions of this approach always cast it as merely an approximation to the continuum problem, but a more
sophisticated approach will seek to determine the relations that a discrete model exactly satisfies.  In particular,
we will ask what is the form of the continuity equation implied by the discretization of Schroediner's equation
on such an irregular mesh? 

Let us assume that Schroedinger's equation has been discretized on a two-dimensional irregular triangular mesh using the 
``cotangent formula'' for the Laplacian \cite{Meyer2002}.  
\begin{equation}
 (L\psi)_i = \frac{1}{A_i} \sum_{j} \omega_{i,j} \left(\psi_j - \psi_i \right),
\end{equation}
where $\omega_{i,j}$ is the sum of two cotangents of angles contained in the two triangles which share the $i$-$j$ edge, and
zero if the points $i$ and $j$ are more remotely located.  Also, $\omega_{i,j} = \omega_{j,i}$ and $A_i$ is the area of
the cell enclosing point $i$.  If we use this Laplacian in the Schroedinger equation, and construct the continuity equation in the usual way,
we find:
\begin{align}
 \frac{\partial \rho_i}{\partial t} \equiv \frac{\partial}{\partial t} \left( A_i \psi^*_i \psi_i \right)
  &= -\frac{i \hbar}{2 m} \sum_j \left[ \psi_i \omega_{i,j} \left(\psi^*_j - \psi^*_i \right) - \psi_i^* \omega_{i,j}\left(\psi_j - \psi_i \right) \right], \notag \\
  &= -\frac{i \hbar}{2 m} \sum_j \omega_{i,j} \left( \psi^*_j \psi_i - \psi^*_i \psi_j \right) .
\end{align}
Thus, the finite-element case has the same structure as the Wannier-Orbital case, with a different current density associated 
with each pair of coupled mesh points.  Observe that we have to take into account the variation of area associated with each
mesh point, to express the continuity relation in terms of the total probability within each cell.

  One potential complication with this formulation is that the coupling elements $\omega_{i,j}$ are not guaranteed to be non-negative,
and negative values are known to occur if oblique triangles are present in the mesh.  This will produce currents that flow in the
``wrong'' direction, but such effects also occur in solutions of the classical diffusion equation\cite{Wardetzky2007}.

\subsection{Relation Between Current Density and the Velocity Operator}

Another procedure which has been used to derive the current density is to identify it with the velocity operator,
as defined by the Heisenberg equation of motion:
\begin{equation} \label{eqn:velocityOp}
  J \leftrightarrow v = (i/\hbar) \left[ H, z \right] .
\end{equation}
For a continuum model, this will produce the same results as (\ref{eqn:wannierJ}) and (\ref{eqn:Jodd_order}) after 
symmetrization of $v$ with respect to its adjoint. (That is, transformation to an anti-commutator superoperator.) 

For discrete formulations, this procedure produces a subtle discrepancy with respect to the discrete continuity equation.
Consider the simple discretization of the effective-mass Hamiltonian and position operator on a uniform mesh of spacing $a$:
\begin{gather*}
 H_{i,j} = \frac{\hbar^2}{2m^* a^2} \left( -\delta_{i-1,j} + 2 \delta_{i,j} - \delta_{i+1,j} \right) , \\
 z_{i,j} = ja \delta_{i,j} .
\end{gather*}
Then,
\begin{equation}
 v_{i,j} = -\frac{i\hbar}{m^*} \left( \frac{\delta_{i+1,j} - \delta_{i-1,j}}{2a} \right).
\end{equation}
Thus, the use of the Heisenberg equation leads to a velocity whose values are coincident with the mesh points and which
is defined by a centered diference.  This is also what would be obtained by application of (\ref{eqn:velocityOp}) followed by 
discretization according to textbook recommendations.

Applying the adjoint symmetrization, we find the explicit form:
\begin{equation}
 \langle v \rangle_i = -\frac{i\hbar}{4m^*a} \left( \psi^*_{i+1} \psi_i  - \psi_i^* \psi_{i+1}  +  \psi_i^* \psi_{i-1} - \psi_{i-1}^* \psi_i \right).
\end{equation}
  The tight-binding approach described above is equally applicable to this case, but
it produces a current density value which is associated with the interval between meshpoints (as the $i+1/2$ index implies),
yielding the explicit expression:
\begin{equation}
 \langle J \rangle_{i+1/2} = \frac{ \hbar}{2im^*a} \left( \psi_{i+1}^* \psi_i - \psi_i^* \psi_{i+1}   \right),
\end{equation} 
which will exactly satisfy the continuity equation (\ref{eqn:discrContinuity1}).
These differing results are related by:
\begin{equation}
\langle v \rangle_i = \frac{1}{2} \left( \langle J \rangle_{i-1/2} + \langle J \rangle_{i+1/2}  \right).
\end{equation}

In a steady-state problem in one dimension, all of the currents discussed above will be equal, and equal to those at any
other location.  In transient situations, however, the central-difference will not exactly satisfy any simple continuity
equation.  Maintaining the central-diference assumption, the net inflow of $\langle v \rangle$ will be:
\begin{align}
 - \nabla \cdot \langle v \rangle_i &\leftrightarrow \frac{1}{2} \left( \langle v \rangle_{i-1} - \langle v \rangle_{i+1} \right) , \notag \\
  &= \frac{1}{4} \left( \langle J \rangle_{i-3/2} + \langle J \rangle_{i-1/2} - \langle J \rangle_{i+1/2} - \langle J \rangle_{i+3/2} \right) , \notag \\
  &= \frac{1}{2} \frac{\partial \rho_i}{\partial t} + \frac{1}{4} \left( \langle J \rangle_{i-3/2} - \langle J \rangle_{i+3/2} \right).
\end{align}
This is clearly not going to lead to any sensible continuity equation, as it invokes more remote current components. Consequently we
must conclude that the Heisenberg velocity cannot be identified with the current density in discrete domains.  

The expected rebuttal to the argument that I have just made is that the discrepancies will disappear as we let the mesh spacing approach zero.
That cannot be taken for granted when the central-difference approximation to the gradient is employed.  Because the central difference 
produces an anomalous value of zero for the maximum spatial Fourier component $k=\pi/a$, the convergence to the continuum is not uniform.
This is the origin of the long-standing problem of anomalous states in envelope-function models of quantum states in heterostructures, and
the remedy to this problem is the use of first-order differences \cite{Frensley2015}.

\section*{Additional References}

\bibitem{Meyer2002} M.~Meyer, {\it et al.}, ``Discrete Differential-Geometry Operators for Triangulated 2-Manifolds,''
{\it Visualization and Mathematics III}, Hege and Polthier, eds., pp.\ 35-57 (Springer, 2002).

\bibitem{Wardetzky2007} M.~Wardetzky, {\it et al.}, ``Discrete Laplace Operators: No Free Lunch,''
{\it Proceedings of the Fifth Eurographics Symposium on Geometry Processing}, Barcelona, Spain, 2007, pp.\ 33--7.

\bibitem{Frensley2015} W.~R. Frensley and R.~Mir,  arXiv 1412.7201 (2015).


\end{document}